# A novel zero-frequency seismic metamaterial

Yi Zeng[1,2], Pai Peng[2], Qiu-Jiao Du[2,*] and Yue-Sheng Wang[1,*]

[1]Department of Mechanics, School of Mechanical Engineering, Tianjin University, Tianjin 300350, China

[2] School of Mathematics and Physics, China University of Geosciences, Wuhan 430074, China

* Electronic mail: qiujiaodu@cug.edu.cn (Q.-J. Du), yswang@tju.edu.cn (Y.-S. Wang).

**Abstract**

A zero-frequency seismic metamaterial (ZFSM) consisting of a three-component seismic metamaterial plate and a half space is proposed to attenuate ultra-low frequency seismic surface waves. The design concept and models are verified firstly by lab-scale experiments on the seismic metamaterial consisting of a two-component seismic metamaterial plate and a half space. Then we calculate the band structures of the one-dimensional and two-dimensional ZFSMs, and evaluate their attenuation ability to Rayleigh waves. A wide band gap and a zero-frequency band gap (ZFBG) can be found as the band structure of the seismic metamaterial is almost equal to the band structure of the seismic metamaterial plate plus the sound cone. It is found that the Rayleigh waves in the ZFSM are deflected and converted into bulk waves. When the number of the unit cells of the ZFSM is sufficient, the transmission distance and deflection angle of the Rayleigh waves in the ZFSM are constant at the same frequency. This discovery is expected to open up the possibility of seismic protection for large nuclear power plants, ancient buildings and metropolitan areas.

# 1 Introduction

Millions of earthquakes occur every year on the planet we live in.[1] According to the statistics of China Earthquake Networks Center, there have been 836 major earthquakes of magnitude 6 and above in the past 100 years (1919-2018), i.e. an average of more than eight times a year. The major casualties caused by Earthquakes are due to the collapse of buildings. The destruction of some important buildings and facilities will bring devastating disasters and irreparable damage, such as nuclear power plants and ancient buildings. The seismic metamaterials (SMs)[2-5] proposed in recent years can prevent these things in a clever way.

SMs with periodically arranged structure have band gaps.[6-9] In the frequency ranges of the band gaps, the seismic waves cannot pass through the SMs. In 2014, an experiment with cylindrical holes arranged periodically verified that this SM can attenuate seismic waves in the band gap.[2] Even though the shape of the holes or the material filled in the holes, the band gaps and the attenuation effect of the SMs on the seismic waves can still be found.[10-12] It has been found that the forest can attenuate the seismic surface waves as the pillars arranged periodically on the half space can open the band gap of the surface waves.[13] Therefore, various pillars placed on the ground have been extensively studied and experimentally verified.[14-17] It is worth mentioning that the use of fractal structures [18] and "rainbow trapping effect" [4] can realize the attenuation of seismic surface waves over a wider range of frequency. In addition, some multi-component SMs have been proposed to attenuate the seismic bulk waves and surface waves.[3,19-25] However, few SMs can attenuate seismic waves below 5 Hz which contains the resonant frequencies of many large buildings and metropolises. It is refreshing that there are also some works [26,27] that consider the effectiveness of SMs in layered soils, which will facilitate the practical applications of SMs. Although SMs clamped by bedrock have zero-frequency band gap (ZFBG), they are only suitable for areas where the embedded depth of bedrock is shallow.[27,28]

In this paper, a zero-frequency SM (ZFSM) consisting of three-component seismic metamaterial plate (SMP) and a half space is proposed to attenuate ultra-low frequency seismic surface waves in a variety of geological situations. We first verify our design method by lab-scale experiments on surface wave transmission of a two-component SM consisting of the two-component SMP and a half space. Then we calculate the band structures of the one-dimensional (1D) and two-dimensional (2D) ZFSMs, check their attenuation ability to Rayleigh waves, and specifically discuss the propagation of Rayleigh waves in the ZFSM in the range of ZFBG. The ZFSM uses traditional building materials and owns wide band gaps and ZFBG, making it a very practical design.

## 2 Method

### 2.1 Numerical simulations

In this paper, we use Solid Mechanics Module of COMSOL Multiphysics to calculate the band structures and the propagation of Rayleigh waves in the SM. For the one-dimensional periodic SM, the reduced wave vector from 0 to $\pi/a$ ($a$ is the period of the SM) is used to obtain the eigenvalues.[29] For the two-dimensional periodic SM, which is highly symmetrical, the wave vector along $\Gamma - X - M - \Gamma$ direction[30] is used to calculate the band structures. For the propagation of Rayleigh waves in the SM, the wave source has a sagittal displacement in *xz* plane on the surface to obtain Rayleigh waves which are often concerned in the study of SMs as they can travel far away and are extremely harmful to buildings.[4,17]

Figure 1(a) shows that the unit cell of the two-component SMP consists of a steel core wrapped with a flexible foam plate. Figure 1(b) shows the unit cell of the two-component SM consisting of the two-component SMP and half space. The band structures of two-component SMP and SM are calculated along $\Gamma - X - M - \Gamma$ direction, which are shown in Figs. 1(c) and 1(d) respectively.

**2.2 Lab-scale experimental setup**

In the experiment, the sample comprising foam-steel composite is zoomed in the upper right corner of Fig. 1(e). The size of the steel core is 140 mm x 40 mm x 40 mm, and the thickness of the flexible foam plate is 5 mm. The size of the glass box is 1.2 m x 0.5 m x 0.5 m, and the material of the substrate is sand. In order to reduce the reflection of the elastic wave by the glass wall, pebbles are placed around the wall.[22] The signal amplifier (label B in Fig. 1(e)) (SP, Inc., YE5874A) amplifies the signal from the signal generator (label A in Fig. 1(e)) (uT, uT8904FRS-DY) to the exciter. The exciter (label C in Fig. 1(e)) (SP, Inc, JZK-50) excites vibrations on one side of the box. The accelerations at points D and E are recorded and analyzed by the computer (label F in Fig. 1(e)).

The accelerations are carried out by using two accelerometers (X&K Tech., XK101S) placed on two symmetrical points (labels D and E in Fig. 1(e)). One (label D) is placed in the free zone where there is no interaction between the SM and the seismic wave, and the other ((label E)) is located behind the SM. Suppose that the acceleration values of points D and E are $A_D$ and $A_E$, respectively. Then the attenuation spectrum of SM on the surface wave is $20 \times \log_{10}(A_E/A_D)$ shown in Fig. 1(f).

## 3 Discussions of models

### 3.1 Experimental verification of numerical model of two-component SM

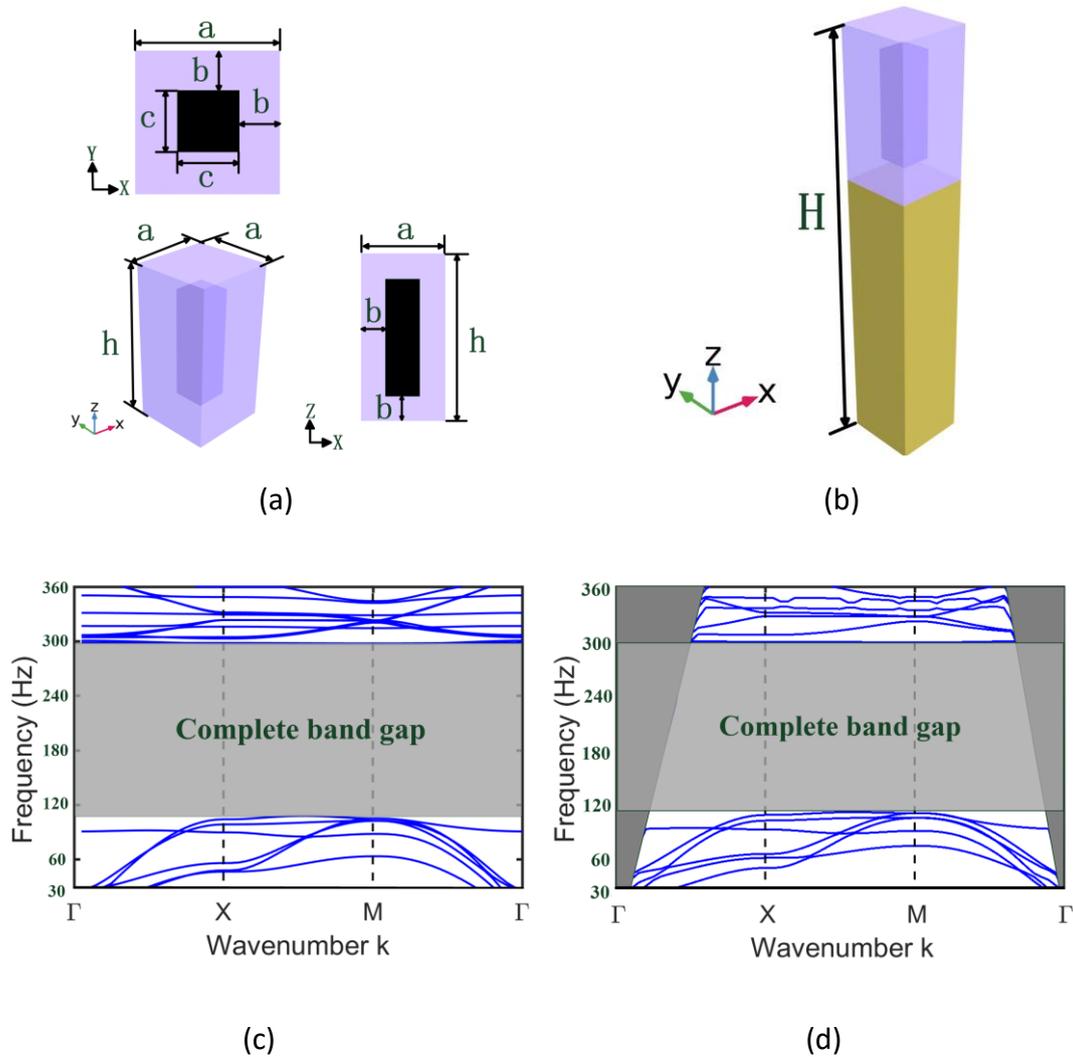

(a)　(b)　(c)　(d)

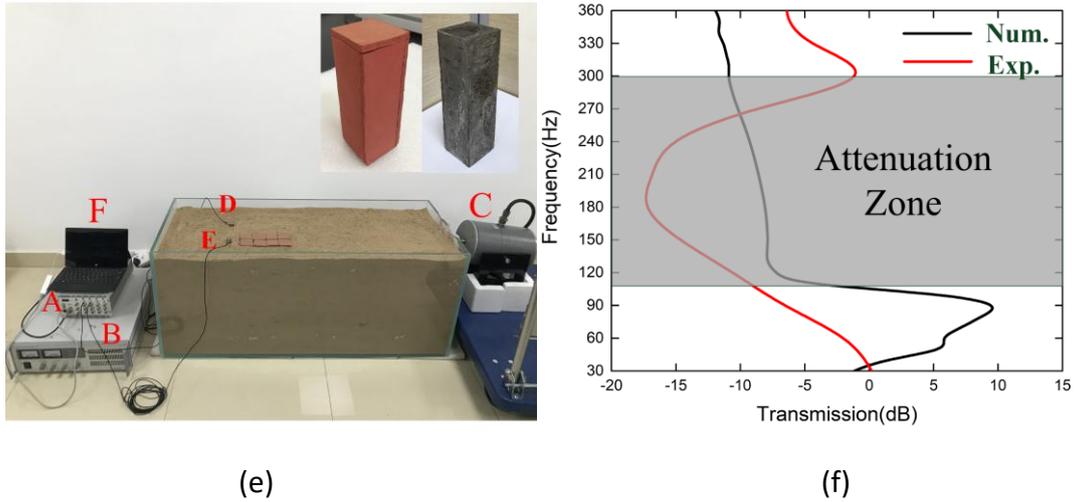

(e)                      (f)

Figure 1 (a) The unit cell of two-component SMP. (b) The unit cell of two-component SM consisting of two-component SMP and half space. Note that the yellow, purple and black areas represent soil, flexible foamed plate and steel core, respectively. Band structures of (c) two-component SMP and (d) two-component SM. (e) Experimental setup and the design of unit cell. (f) Experimental and numerical transmission spectrums for supercell of eight unit cells (four rows) of two-component SM.

In order to verify that new design of SMs can attenuate seismic surface waves, we perform lab-scale experiment for a supercell of eight unit cells (four rows) of two-component SM to measure the transmission spectrum. And the geometric parameters are: $a = 5$ cm, $b = 1$ cm, $c = 4$ cm, $h = 15$ cm and $H = 20a$. The material parameters [31] are illustrated in Table 1. All parameters used in the simulation are consistent with those in the lab-scale experiment.

Comparing the two band structures, we can find that when the SMP is placed on the half space, the original band is basically unchanged, but the sound cone (dark gray area in Fig. 1(d)) is added due to the existence of the half space. So we can not only keep the complete band gap (CBG) of the SMP itself in Fig. 1(d). It is found that the addition of the substrate does not affect the original band structure of SMP. That is to say, the band structure of the SM composed of the two-component SMP and a half space is almost equal to that of the two-component SMP plus the sound cone. Therefore, the wide band gap of the SMP itself is preserved. At 110 Hz – 300 Hz, we can verify that the SM can attenuate the surface waves well by lab-scale experiments and simulated calculations of the supercell, which corresponds to the band gap in Fig. 1(d). The geometric model, material parameters, and position of the supercell and substrate used in the simulation calculations are consistent with the lab-scale experiments. In addition, the boundary formed by the glass wall is replaced by a low reflection boundary to reduce reflected waves.

Table 1. Material parameters of the unit cell and the half space

| Material | Density (kg/m$^3$) | Young's modulus (Pa) | Poisson's ratio |
| --- | --- | --- | --- |
| Rubber | 1300 | $1.2 \times 10^5$ | 0.47 |
| Concrete | 2500 | $4 \times 10^{10}$ | 0.3 |
| Steel | 7784 | $2.07 \times 10^{11}$ | 0.3 |
| foamed plate | 1053 | $1.6 \times 10^5$ | 0.39 |
| Soil | 1800 | $2 \times 10^7$ | 0.3 |

**3.2 Discussions of ZFSM**

From the above simulations and lab-scale experiment of the SM consisting of the two-component SMP and half space, it can be seen that the band structure of the SM is almost equal to the band structure of the SMP plus the sound cone. And this SM can effectively attenuate the seismic surface waves to protect the buildings. Next, we will use FEM calculations to discuss the band structure and attenuation effect of the ZFSM consisting of the three-component SMP and a half space on the seismic surface waves. The ZFBG of the ZFSM is discussed under the one-dimensional (1D) and two-dimensional (2D) periodic models.

**3.2.1 Model of 1D ZFSM**

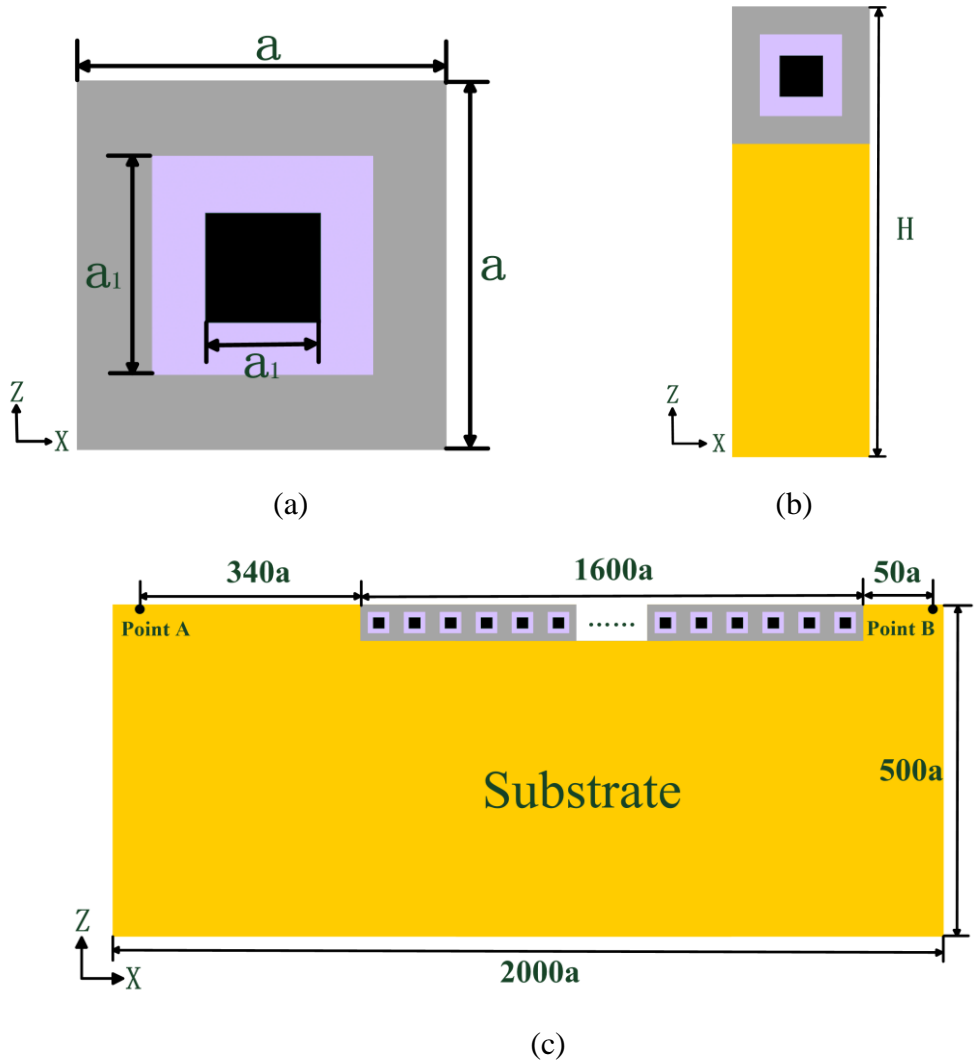

Figure 2 (a) The unit cell of the 1D three-component SMP. (b) The unit cell of 1D ZFSM. (c) A schematic diagram of a system for calculating transmission spectra. Note that the yellow, gray, purple and black areas represent soil, concrete, rubber and steel core, respectively.

Figure 2(a) shows a unit cell of 1D three-component SMP, which is set to a Bloch periodic boundary condition in the *x* direction and a free boundary condition in the *z* direction. Figure 2(b) shows a unit cell of the 1D ZFSM consists of a unit cell of 1D three-component SMP and a half space. It is set to the Bloch period boundary condition in the *x* direction, the upper boundary is a free boundary, and the lower boundary is a fixed boundary. And the geometric parameters are: $a = 1.5$ m, $a_1 = 0.6a$, $a_2 = 0.5a$, $H = 20a$. The material parameters are illustrated in Table 1.[11] Figure 2(c) shows a schematic diagram of the system used to calculate the transmission spectrum. Since the ZFBG of the ZFSM is mainly studied and discussed in this paper, and the wavelength of the seismic surface wave is large in the range of ZFBG, the values of the geometric parameters involved here are all large, especially the depth ($500a$) of the substrate, the distance ($340a$) from the source (point A in Fig. 2(c)) to the ZFSM and the number (1600) of the unit cells of the ZFSM. Point B (Fig. 2(c)) is the signal collection point. In addition, the bottom and left and right sides of the substrate are set as low reflection boundaries to reduce reflection. Suppose that the acceleration at point B located after the SMP is $A_1$, and the acceleration at point B without the SMP is $A_0$. Then the transmission function is defined as $20 \times \log_{10}(A_1/A_0)$.

### 3.2.2 Band structure and transmission spectrum of 1D ZFSM

For a 1D three-component SMP based on local resonance, a band gap can easily occur. As shown in Fig. 3(a), a narrow band gap appears in the range of 10.0 – 14.0 Hz. Even if the unit cells in Fig. 2(a) are periodically arranged in two dimensions, the complete band gap is very narrow.[20] It is conceivable that such a narrow band gap cannot meet our needs. However when we place this SMP on a half space (i.e. ZFSM), we are surprised to find two wide band gaps (0 – 9.0 Hz and 9.4 – 22.0 Hz) for surface waves. And one of them is a ZFBG. By comparing the band structures in Figs. 3(a) and 3(b), we can easily find that the band structure of the 1D ZFSM consisting of the 1D three-component SMP and a half space is almost equal to the band structure of the 1D three-component SMP plus the sound cone. Therefore, the zero-frequency band gap and the wide band gap are generated shown in Fig. 3(b).

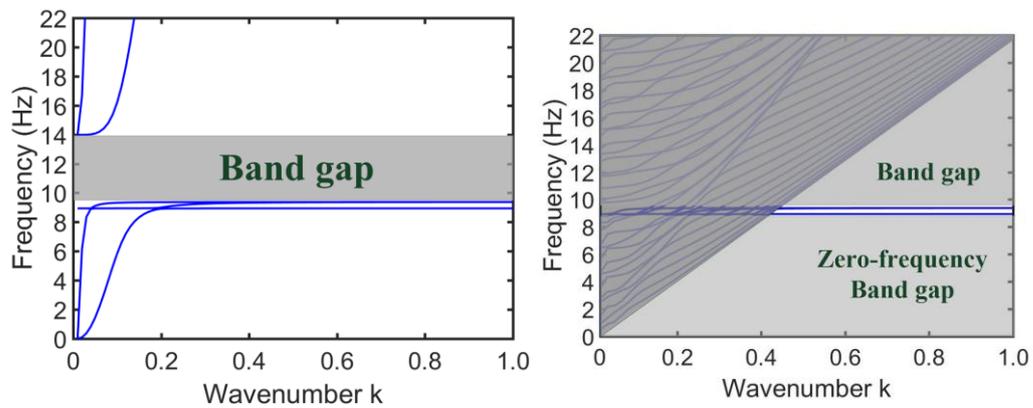

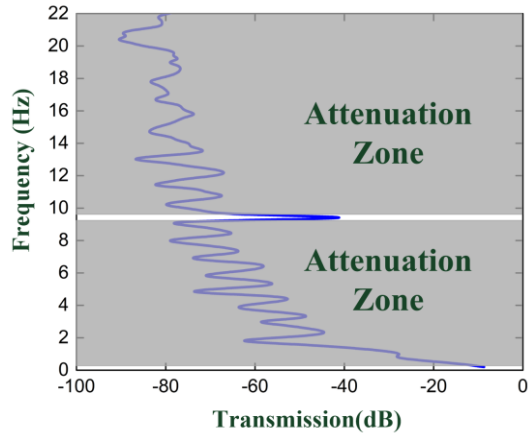

Figure 3 (a) Band structure of 1D three-component SMP. (b) Band structure of 1D ZFSM. (c) Transmission spectrum of the system in Fig. 2(c).

Through the transmission spectrum of the ZFSM, we can also see the existence of these two wide band gaps. In the range of the ZFBG, we see that the seismic surface waves are well attenuated in the range of 0.5 – 9.0 Hz. However, for the seismic surface waves close to zero frequency, the attenuation effect needs to be improved. Because the number of unit cells in the proposed ZFSM is infinite because of the periodic boundary condition when calculating the band structure, while the number of the unit cells in the ZFSM is finite (1600 unit cells) when calculating the transmission spectrum. Therefore, we hypothesize that the number of the unit cells of the ZFSM and the lower limit of the attenuation zone in the transmission spectrum are closely related.

### 3.2.3 Discussion on ZFBG of 1D ZFSM

To discover the relationship between the number of the unit cells of the ZFSM and the lower limit of the attenuation zone in the transmission spectrum, we simulate and plot the displacements of surface waves of the system varying with the frequency by calculating the transmission spectrum. As shown in Fig. 4(a), we use the ZFSM of 1600 unit cells placed from 0 m to 2400 m for calculation. As the frequency decreases from 5.0 Hz to around 0.5 Hz, the transmission distance of the Rayleigh waves increases exponentially in the ZFSM. When the frequency is less than 0.5 Hz, Rayleigh waves pass through the 1600 unit cells of ZFSM. We can speculate that if the number of unit cells in the ZFSM increases, the propagation of the lower-frequency Rayleigh waves can be attenuated. Physically speaking, this ZFSM is a novel SM with ZFBG. For practical applications, it is also very convenient to select the number of unit cells according to the frequency range that we consider. We have studied the transmission of Rayleigh waves in the ZFSM at 2.2 Hz to demonstrate the mechanism of the ZFSM. As shown in Fig. 4(c), when the Rayleigh waves enter the ZFSM, they are deflected at a certain angle and converted into bulk waves.

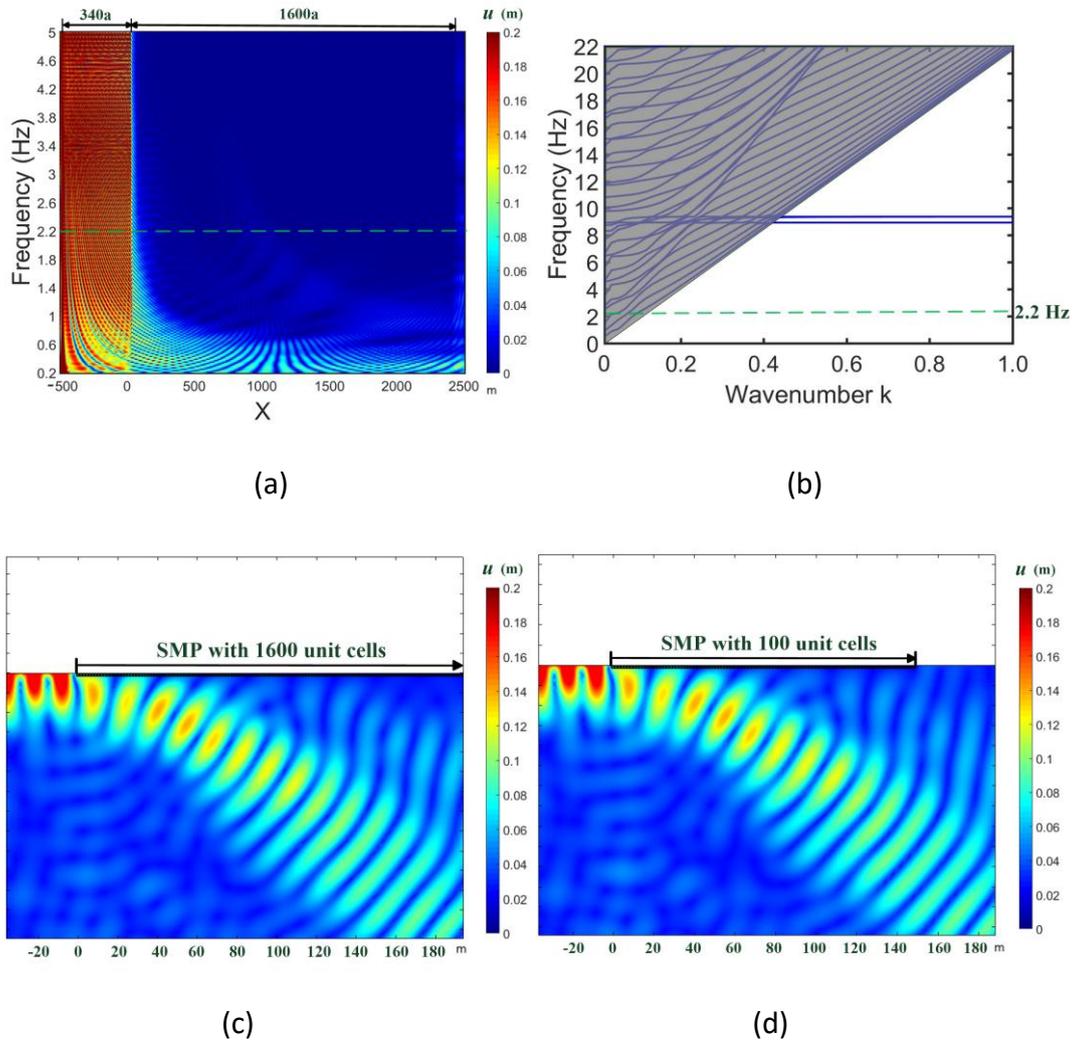

(a)                          (b)

(c)                          (d)

Figure 4 (a) The displacements of surface waves of the system varying with the frequency by calculating the transmission spectrum. (b) ZFSM's band structure and a dotted line with a frequency of 2.2 Hz. (c) The propagation of Rayleigh waves in ZFSM with 1600 unit cells at 2.2 Hz. (d) The propagation of Rayleigh waves in ZFSM with 100 unit cells at 2.2 Hz. The colors indicate the total displacement ($u$) ranging from 0 to maximum.

To explain this phenomenon, we draw a dotted line with the frequency of 2.2 Hz on the band structure of the ZFSM shown in Fig. 4(b). At 2.2 Hz, there is no surface mode in the ZFSM. However, there are a large number of bulk wave modes outside the sound cone. That is to say, at 2.2 Hz, only the bulk wave modes exist in the ZFSM. Therefore, Rayleigh waves will be deflected and converted into bulk waves as shown in Fig. 4(c). This principle is also used to achieve low-consumption energy transmission and other functions in topological metamaterials and has been validated in experiments.[32,33]

We also use the ZFSM of 100 unit cells placed from 0 m to 150 m with for the calculation of the transmission spectrum in Fig. 4(d). We find that the deflection of Rayleigh waves is almost the same as the ZFSM with 1600 unit cells. When the Rayleigh waves propagate to the position of $X = 60$ m (presumably at the position of the 40th unit cell), most of the energy has turned from the surface to the bulk below the surface. This also shows that when the number of the unit cells of the ZFSM is sufficient, the transmission distance and deflection angle of the Rayleigh waves in the ZFSM are constant at the same frequency. In practical applications, the method of arranging the number of the unit cells of the ZFSM according to the specified frequency range is effective.

In order to investigate the influence of the material parameters of the substrate on the ZFBG, we consider the effect of the Young's modulus (Fig. 5(a)) and mass density (Fig. 5(b)) of the substrate on the upper frequency bounds of the ZFBG for the 1D ZFSM. When the Young's modulus of the substrate increases from 1 kPa to 3.4 MPa, the width of the ZFBG is enlarged as its upper boundary gradually increases. In this range of the Young's modulus, the upper boundary of the ZFBG is equal to the frequency value of the apex of the sound cone, which is the value of the fast transverse wave threshold of the substrate at the wavenumber $k = \pi/a$.[29] This value gradually increases as the Young's modulus increases. When the Young's modulus of the substrate increases from 3.4 MPa to 50 MPa, the upper boundary (9.0 Hz) of the ZFBG does not change. In this range of the Young's modulus, the upper boundary of the ZFBG is determined by the band of the local resonance of the SMP, which is invariable. And the Young's modulus of the soil (20 MPa) used in this paper is within this range. When the Young's modulus of the substrate increases from 50 MPa to 500 MPa, the ZFBG disappears as the upper boundary drops sharply. When the Young's modulus of the substrate increases from 500 MPa to 100 GPa, the ZFBG does not reappear. The increase of the Young's modulus of the substrate causes the increase of the slope of the fast transverse wave threshold, and consequently the sound cone cannot cover the band of the SMP at low frequencies. Therefore, the upper boundary of the ZFBG rapidly drops and the ZFBG disappears.

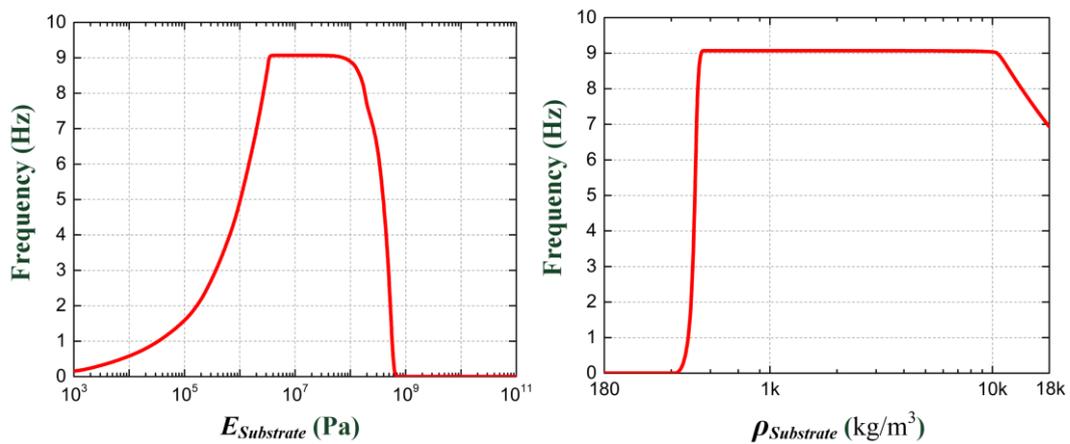

Figure 5 The effect of (a) Young's modulus and (b) mass density of substrate on the upper frequency bounds of the ZFBG for the 1D ZFSM

There is no ZFBG when the mass density of the substrate is less than 470 kg/m$^3$. The low mass density of the substrate leads to the high slope of the fast transverse wave threshold, so that the sound cone cannot cover the band of the SMP at low frequencies, which results in the disappearance of the ZFBG. When the mass density of the substrate is between 470 and 10260 kg/m$^3$, the upper boundary (9.0 Hz) of the ZFBG does not change. This is consistent with the reason why the ZFBG in the previous paragraph does not change. The mass density of the soil (1800 kg/m$^3$) used in this paper is within this range. When the mass density of the substrate increases from 10260 kg/m$^3$ to 18000 kg/m$^3$, the width of the ZFBG gradually decreases. Within this range of the mass density, the upper boundary of ZFBG is equal to the frequency value of the apex of the sound cone. When the mass density of substrate increases, the value gradually decreases. Therefore, the upper boundary of ZFBG is gradually reduced.

From the effect of the Young's modulus and mass density of the substrate on the upper frequency bounds of the ZFBG for the 1D ZFSM, we can find that the charge of the material parameters of the substrate only affects the slope of the fast transverse wave threshold. When the slope is too large, the sound cone cannot cover the band of the SMP at low frequencies, resulting in the disappearance of the ZFBG. When the slope is too small, the cone will cover the band of the local resonance of the SMP, and the width of the ZFBG will decrease. When the slope is appropriate, the width of the ZFBG may reach its largest value (9.0Hz). That is to say, the band structure of the SM composed of the SMP and a half space is almost equal to that of the SMP plus the sound cone. It is noteworthy that this conclusion is incorrect when the heavy core of the unit cell of the SMP connects the substrate.[10,11]

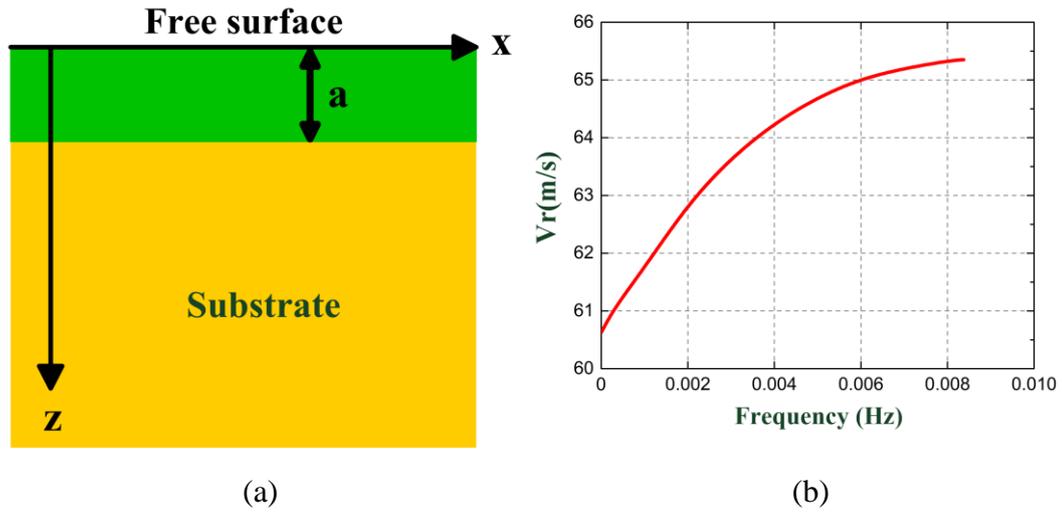

(a)          (b)

Figure 6 (a) Effective two-layered medium of 1D ZFSM, (b) Rayleigh waves dispersion curve of the two-layered media.

The above 1D ZFSM can be regarded as a two-layered medium as shown in Fig. 6(a). The green part is the effective medium of the SMP. Its upper boundary is the free boundary, and the thickness is $a = 1.5$ m. In the low frequency region (0 - 5 Hz) away from local resonance, the effective mass density of the SMP

$$\rho_e = \phi_1 \rho_1 + \phi_2 \rho_2 + \phi_3 \rho_3$$

where $\phi_1$, $\phi_2$ and $\phi_3$ are the filling rates of concrete, rubber and steel core, respectively, satisfying $\phi_1 + \phi_2 + \phi_3 = 1$. And $\rho_1$, $\rho_2$ and $\rho_3$ are the mass densities of concrete, rubber and steel core, respectively.[34,35] If the resonance frequency is ignored, the bulk modulus and shear modulus of the SMP and concrete (substrate of the SMP) are very close.[35,36] Therefore, the material parameters of the SMP are set as follows: mass density $\rho_e = 4588.36$ kg/m$^3$, Young's modulus $E_e = 40$ GPa, Poisson's ratio $\upsilon_e = 0.3$. The substrate shown in Fig. 6(a) is a half space in which the material is soil.

For the layered medium shown in Fig. 6(a), the dispersion curve of the Rayleigh wave can be calculated by the fast scalar method.[37] As shown in Fig. 6(b), the dispersion curve of the layered medium only exists in a single waveguide mode and exhibits inverse dispersion[38]. In the range of 0 - 0.0084 Hz, the speed of the Rayleigh wave gradually increases with increasing frequency. It is worth noting that the Rayleigh wave has a cutoff frequency (0.0084 Hz). That is to say, in the low frequency region where local resonance does also not exist, there is no Rayleigh wave when the frequency is higher than 0.0084 Hz in this effective two-layered medium. This also verifies the correctness of the ZFBG because 0.0084Hz is close to zero.

**3.2.4 Model of 2D ZFSM**

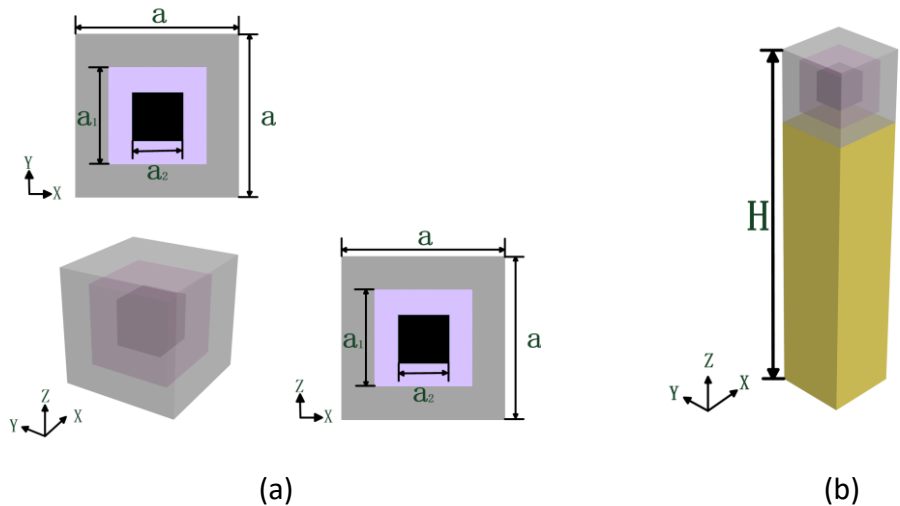

(a)          (b)

Figure 7 (a) The unit cell of the 2D three-component SMP. (b) The unit cell of 2D ZFSM. Note that the yellow, gray, purple and black areas represent soil, concrete, rubber and steel core, respectively.

In order to verify that the above the ZFSM still has a ZFBG in the three-dimensional space, we have studied the 2D ZFSM as shown in Fig. 7(b). Similarly, we also calculate the band structure of the unit cell of the 2D three-component SMP shown in

Fig. 7(a). And the unit cell of the 2D ZFSM consists of a unit cell of the 2D three-component SMP and a half space. The geometric parameters ($a$, $a_1$, $a_2$ and $H$) and the material parameters of the four materials (concrete, rubber, steel and soil) are consistent with those of the 1D ZFSM. We set the Bloch period boundary conditions on the four vertical boundaries of the 2D SMP; and the upper and lower boundaries are free. The Bloch period boundary condition is set on the four vertical boundaries of the 2D ZFSM; the lower boundary is set as the fixed boundary condition, and the upper boundary is set as the free boundary condition. The system used for transmission spectrum calculation is similar to Fig. 2(c), so it is not illustrated here.[17]

**3.2.5 Band structure and transmission spectrum of 2D ZFSM**

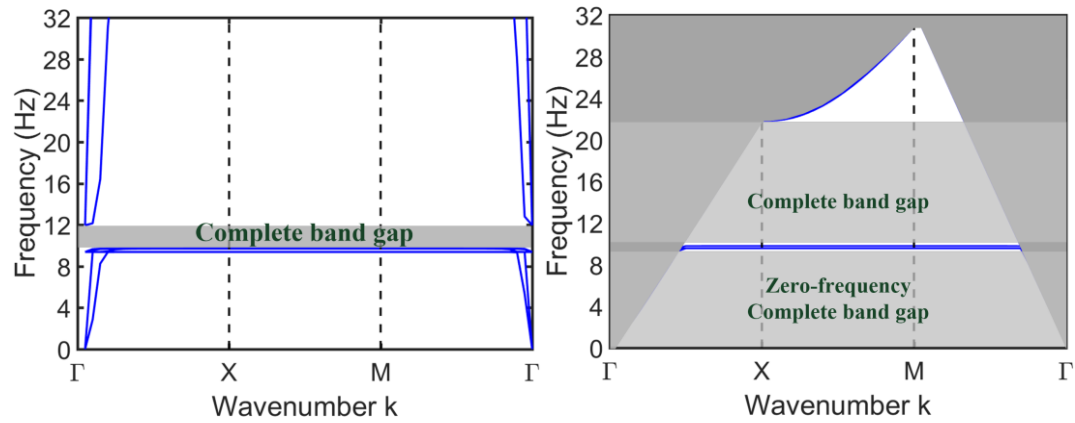

(a)                                     (b)

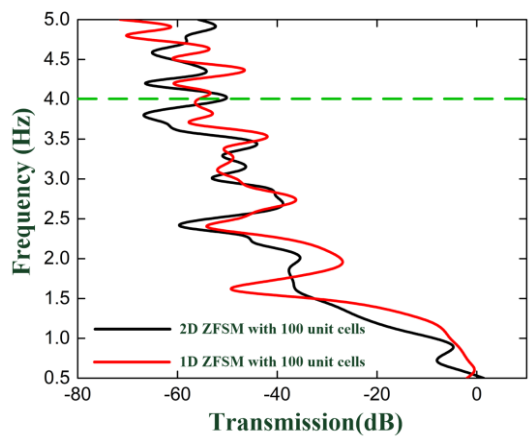

(c)

Figure 8 (a) Band structure of the 2D three-component SMP. (b) Band structure of the 2D ZFSM. (c) Transmission spectra of 1D and 2D ZFSMs with 100 unit cells in the frequency range of 0.5 Hz - 5.0 Hz.

Metamaterials based on the local resonance is the subject of discussion. Although they can obtain CBGs at low frequencies, the width of the CBGs needs to be expanded.[39-41] As shown in Fig. 8(a), the CBG of the 2D three-component SMP is very narrow and only exists around 10.0 – 12.0 Hz. Even if the unit cells of such three-component SMP are three-dimensionally periodically arranged, the width of the CBG is also narrow.[21] However, the 2D ZFSM has two wide CBG for surface waves, and one of them is a zero-frequency CBG. By comparing the band structures in Figs. 8(a) and 8(b), we can easily find that the band structure of the ZFSM consisting of the 2D three-component SMP and a half space is almost equal to the band structure of the 2D three-component SMP plus the sound cone. This is also the reason for the wide zero-frequency CBG.

We calculate the transmission spectrums of Rayleigh waves in the 1D and 2D ZFSMs (along $\Gamma X$ direction) with 100 unit cells at 0.5 – 5.0 Hz which is within the ZFBG. The attenuation effect of the two kinds of ZFSM is almost the same, and the 2D ZFSM is slightly better than the 1D ZFSM at low frequencies (0.5 – 4.0 Hz). This shows that the 2D ZFSM can also attenuate ultra-low frequency surface waves in the three-dimensional space.

**3.2.6 Discussion on ZFBG of 2D ZFSM**

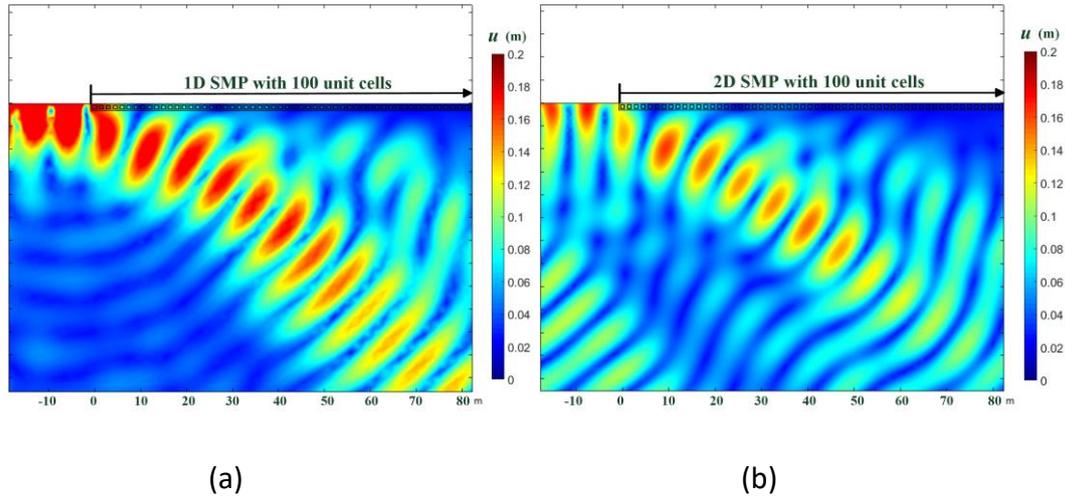

(a)                          (b)

Figure 9 The propagations of Rayleigh waves in (a) 1D ZFSM and (b) 2D ZFSM (*xz* section diagram) with 100 unit cells at 4.0 Hz. The colors indicate the total displacement (*u*) ranging from 0 to maximum.

As shown in Fig. 9, by comparing the propagations of Rayleigh waves in the 1D and 2D ZFSM (xz section diagram) with 100 unit cells at 4.0 Hz, the Rayleigh waves in the 2D ZFSM also convert from surface waves to bulk waves, and the deflection angle of the Rayleigh wave is almost the same as that of the 1D ZFSM. This shows that the 2D ZFSM can attenuate surface waves in the frequency range of its ZFBG to protect the target building. Its mechanism is consistent with that of the 1D ZFSM.

**4 Conclusions**

In summary, through the lab-scale experiment and numerical simulation of the ZFSM consisting of the two-component SMP and a half space, we demonstrate this new design methodology to create SMs to attenuate seismic surface waves. Then we propose a three-component ZFSM consisting of the three-component SMP and a half space. By investigating the influence of the material parameters of the substrate on the 1D ZFBG, we find that the band structure of the SM composed of the SMP and a half space is almost equal to that of the SMP plus the sound cone. By analyzing the propagation of Rayleigh waves in the 1D and 2D ZFSMs in the frequency range of the ZFBG, we find that Rayleigh waves are deflected at a certain angle and converted into bulk waves. When the number of the unit cells of the ZFSM is sufficient, the transmission distance and deflection angle of the Rayleigh waves in the ZFSM are constant at the same frequency. All the phenomena demonstrate that the ZFSM can attenuate surface waves in the frequency range of its ZFBG to protect the target building. This ZFSM, which is simple and easy to implement, is expected to protect large nuclear power plants, ancient buildings and metropolises from damage by seismic waves. This paper not only demonstrates the application prospects of ZFSM, but also provides design guidance for surface wave researchers.

**Acknowledgments**

We thank Key Program of the National Natural Science Foundation of China (NNSFC) under Grant No. 41830537, the National Programme on Global Change and Air-Sea Interaction (GASI-GEOGE-02), and the National Natural Science Foundation of China (NNSFC) under Grant Nos. 11604307 and 11702017.

**References**

[1] R. K. Reitherman, in *Earthquakes and engineers: an international history*, 2012 (American Society of Civil Engineers).
[2] S. Brûlé, E. Javelaud, S. Enoch, and S. Guenneau, Physical review letters **112,** 133901 (2014).
[3] G. Finocchio, O. Casablanca, G. Ricciardi, U. Alibrandi, F. Garescì, M. Chiappini, and B. Azzerboni, Applied Physics Letters **104,** 191903 (2014).
[4] A. Colombi, D. Colquitt, P. Roux, S. Guenneau, and R. V. Craster, Scientific reports **6,** 27717 (2016).
[5] S. Brûlé, S. Enoch, and S. Guenneau, "Emergence of seismic metamaterials: current state and future perspectives", J. Appl. Phys. preprint arXiv:1712.09115 (2017).
[6] F. Meseguer, M. Holgado, D. Caballero, N. Benaches, J. Sánchez-Dehesa, C. López, and J.

| | Llinares, Physical Review B **59,** 12169 (1999). |
|---|---|
| 7 | V. Laude, M. Wilm, S. Benchabane, and A. Khelif, Physical Review E **71,** 036607 (2005). |
| 8 | T.-T. Wu, L.-C. Wu, and Z.-G. Huang, Journal of Applied Physics **97,** 094916 (2005). |
| 9 | A. Khelif, B. Aoubiza, S. Mohammadi, A. Adibi, and V. Laude, Physical Review E **74,** 046610 (2006). |
| 10 | M. Miniaci, A. Krushynska, F. Bosia, and N. M. Pugno, New Journal of Physics **18,** 083041 (2016). |
| 11 | Q. Du, Y. Zeng, G. Huang, and H. Yang, AIP Advances **7,** 075015 (2017). |
| 12 | X. Pu and Z. Shi, Soil Dynamics and Earthquake Engineering **98,** 67 (2017). |
| 13 | A. Colombi, P. Roux, S. Guenneau, P. Gueguen, and R. V. Craster, Scientific reports **6,** 19238 (2016). |
| 14 | Y.-f. Liu, J.-k. Huang, Y.-g. Li, and Z.-f. Shi, Construction and Building Materials **199,** 737 (2019). |
| 15 | P. Roux, D. Bindi, T. Boxberger, A. Colombi, F. Cotton, I. Douste‐Bacque, S. Garambois, P. Gueguen, G. Hillers, and D. Hollis, Seismological Research Letters **89,** 582 (2018). |
| 16 | D. Colquitt, A. Colombi, R. Craster, P. Roux, and S. Guenneau, Journal of the Mechanics and Physics of Solids **99,** 379 (2017). |
| 17 | Y. Zeng, Y. Xu, K. Deng, Z. Zeng, H. Yang, M. Muzamil, and Q. Du, Journal of Applied Physics **123,** 214901 (2018). |
| 18 | Q. Du, Y. Zeng, Y. Xu, H. Yang, and Z. Zeng, Journal of Physics D: Applied Physics **51,** 105104 (2018). |
| 19 | O. Casablanca, G. Ventura, F. Garescì, B. Azzerboni, B. Chiaia, M. Chiappini, and G. Finocchio, Journal of Applied Physics **123,** 174903 (2018). |
| 20 | Y. Yan, A. Laskar, Z. Cheng, F. Menq, Y. Tang, Y. Mo, and Z. Shi, Journal of Applied Physics **116,** 044908 (2014). |
| 21 | Y. Yan, Z. Cheng, F. Menq, Y. L. Mo, Y. Tang, and Z. Shi, Smart Material Structures **24** (2015). |
| 22 | S. Krödel, N. Thomé, and C. Daraio, Extreme Mechanics Letters **4,** 111 (2015). |
| 23 | Z. Cheng and Z. Shi, Earthquake Engineering & Structural Dynamics **47,** 925 (2018). |
| 24 | N. Aravantinos-Zafiris and M. Sigalas, Journal of Applied Physics **118,** 064901 (2015). |
| 25 | A. Palermo, S. Krödel, A. Marzani, and C. Daraio, Scientific reports **6,** 39356 (2016). |
| 26 | X. Pu and Z. Shi, Construction and Building Materials **180,** 177 (2018). |
| 27 | Y. Chen, Q. Feng, F. Scarpa, L. Zuo, and X. Zhuang, Materials & Design**,** 107813 (2019). |
| 28 | Y. Achaoui, T. Antonakakis, S. Brûlé, R. V. Craster, S. Enoch, and S. Guenneau, New Journal of Physics **19,** 063022 (2017). |
| 29 | B. Graczykowski, F. Alzina, J. Gomis-Bresco, and C. Sotomayor Torres, Journal of Applied Physics **119,** 025308 (2016). |
| 30 | F. Maurin, C. Claeys, E. Deckers, and W. Desmet, International Journal of Solids and Structures (2017). |
| 31 | Y. Zeng, Y. Xu, K. Deng, P. Peng, H. Yang, M. Muzamil, and Q. Du, Journal of Applied Physics **125,** 224901 (2019). |
| 32 | J. Lu, C. Qiu, W. Deng, X. Huang, F. Li, F. Zhang, S. Chen, and Z. Liu, Physical review letters **120,** 116802 (2018). |
| 33 | X. Fan, C. Qiu, Y. Shen, H. He, M. Xiao, M. Ke, and Z. Liu, Physical Review Letters **122,** 136802 (2019). |
| 34 | Y. Ding, Z. Liu, C. Qiu, and J. Shi, Physical review letters **99,** 093904 (2007). |

[35] Y. Ding, "Double Negative Metamaterials for Elastic Waves", Thesis, Wuhan University, 2009. (in Chinese)
[36] X.-N. Liu, G.-K. Hu, G.-L. Huang, and C.-T. Sun, Applied Physics Letters **98,** 251907 (2011).
[37] B. Zhang, M. Yu, C.-Q. Lan, and W. Xiong, The Journal of the Acoustical Society of America **100,** 3527 (1996).
[38] T. Yang, "Forward Modeling of Zigzag Dispersion and Pavement Systems Dispersion Curves", Thesis, Central South University 2004. (in Chinese)
[39] Z. Liu, X. Zhang, Y. Mao, Y. Zhu, Z. Yang, C. Chan, and P. Sheng, Science **289,** 1734 (2000).
[40] P. Sheng, X. X. Zhang, Z. Liu, and C. T. Chan, Physica B: Condensed Matter **338,** 201 (2003).
[41] Y. Achaoui, A. Khelif, S. Benchabane, L. Robert, and V. Laude, Physical Review B **83,** 104201 (2011).